\documentclass[aps,prl,twocolumn,superscriptaddress,showpacs]{revtex4}
\usepackage{hyperref}

\usepackage{graphicx}

\usepackage{amsmath}
\usepackage{epsfig}

\begin{document}
\title{Experimental continuous variable cloning of partial quantum information}

\author{Metin Sabuncu}
\affiliation{Department of Physics,
Technical University of Denmark, 2800 Kongens Lyngby, Denmark}
\affiliation{Institut f\"{u}r Optik, Information und Photonik,  Max-Planck Forschungsgruppe, Universit\"{a}t Erlangen-N\"{u}rnberg, G\"{u}nther-Scharowsky str. 1, 91058, Erlangen, Germany}
\email{msabuncu@optik.uni-erlangen.de}
\author{Gerd Leuchs}
\affiliation{Institut f\"{u}r Optik, Information und Photonik,  Max-Planck Forschungsgruppe, Universit\"{a}t Erlangen-N\"{u}rnberg, G\"{u}nther-Scharowsky str. 1, 91058, Erlangen, Germany}
\author{Ulrik L. Andersen}
\affiliation{Department of Physics,
Technical University of Denmark, 2800 Kongens Lyngby, Denmark}
\affiliation{Institut f\"{u}r Optik, Information und Photonik,  Max-Planck Forschungsgruppe, Universit\"{a}t Erlangen-N\"{u}rnberg, G\"{u}nther-Scharowsky str. 1, 91058, Erlangen, Germany}

\date{\today}

\begin{abstract}
The fidelity of a quantum transformation is strongly linked with the prior partial information of the state to be transformed. We illustrate this interesting point by proposing and demonstrating the superior cloning of coherent states with prior partial information. More specifically, we propose two simple transformations that under the Gaussian assumption optimally clone symmetric Gaussian distributions of coherent states as well as coherent states with known phases. Furthermore, we implement for the first time near-optimal state-dependent cloning schemes relying on simple linear optics and feedforward.
\end{abstract}


\maketitle
The ignorance about a given quantum state is what makes quantum protocols difficult to execute in practice or even impossible in principle. For example, high efficiency and deterministic teleportation of a quantum state with no prior information is only possible in the unrealistic limit of perfect entanglement. Furthermore, it is well known that perfect cloning of an arbitrary quantum state is impossible as formulated in the no-cloning theorem\cite{zurek.nature,dieks}.  Luckily, in all practical quantum communication or computation schemes we are not completely ignorant about the set of possible input states which in turn greatly facilitates the execution of these protocols: The more prior information one has about the input alphabet the less resources are needed for the process. 



An interesting example demonstrating the influence of partial quantum information is cloning. For example, the optimised continuous variable (cv) cloner of an arbitrary state (also known as the cv universal cloner) yields a cloning fidelity of 1/2 corresponding to a standard classical distributor \cite{braunst}. However, if the input states are a-priori known to be coherent states (but with unknown amplitude and phase), the fidelity increases to 2/3 \cite{cerf,lindblad}. By further limiting the number of possible input states the fidelity increases even further as theoretically analysed in ref. \cite{cochrane04.pra} for a symmetric Gaussian distribution of coherent states, in refs. \cite{cochrane04.pra,namiki06.xxx,dong07.pra} for coherent states with known phase and in ref. \cite{sacchi07.pra} for phase covariant cloning where the mean amplitude is fixed but the phase random. Cloning of displaced thermal states and squeezed states has also been theoretically analyzed \cite{olivares06.pra}. Despite this high interest in cloning of partial cv quantum information, there has been no experimental demonstrations. Experimental studies have been entirely devoted to cloning of qubits with partial information, e.g. phase covariant cloning \cite{du, demartini, cernoch}.


In this Letter we investigate, theoretically and experimentally, the optimization of a continuous variable quantum cloning machine with respect to two different coherent state alphabets using a simple setup based entirely on linear optics, homodyne detection and feedforward. In particular we propose and experimentally realise an optimal Gaussian cloning machine tailored to clone a symmetric Gaussian alphabet of coherent states as well as coherent states with known phases. 
In addition, we prove the optimality of the latter scheme and find that a fidelity as large as 96.1\% can in principle be achieved. 
         
A common measure of the quality of a cloning operation is the fidelity which is defined as follows. Consider the protocol where the coherent state, $|\alpha\rangle$, to be cloned is chosen from an ensemble defined by $\{p(\alpha),|\alpha\rangle\}$ where $p(\alpha)$ denotes the probability that the state $|\alpha\rangle$ was chosen. This state undergoes a cloning transformation denoted by $\Gamma (\alpha)$. The overlap $\langle\alpha |\Gamma(\alpha)|\alpha\rangle$ then quantifies the quality of cloning a specific member,  $|\alpha\rangle$, of the alphabet. The average fidelity of the cloning action thus reads
\begin{equation}
\bar F= \int{p(\alpha)\langle\alpha |\Gamma(\alpha)|\alpha\rangle}d^2\alpha
\label{fidelity}
\end{equation}
Using the fidelity as a measure, the cloning transformation is optimal when this expression is maximized. Such maximization normally yields a non-Gaussian solution, that is, the optimal map $\Gamma$ is non-Gaussian. However, since the Gaussian cloning transformation is known to be near optimal, we will mainly focus on such maps. 

\begin{figure}[h]
	\centering
		\includegraphics[width=8cm]{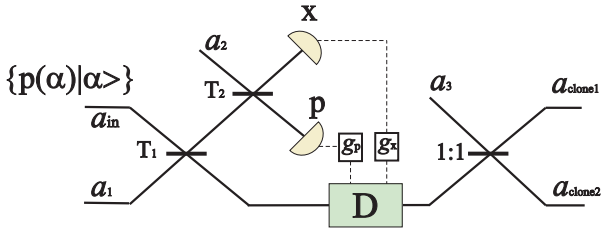}\caption{Schematic of the proposed 1$\rightarrow$2 cloning protocol. The signal, $a_{in}$, is reflected of a beam splitter with transmittance $T_1$ and detected using a beamsplitter with transmittance $T_2$ and two homodyne detectors measuring the amplitude, $x$, and phase, $p$, quadratures. The measurement outcomes are scaled with the gains $g_x$ and $g_p$ and used to displace the transmitted signal. The displaced state is subsequently split on a symmetric beam splitter, thus producing two clones denoted by $a_{clone1}$ and $a_{clone2}$. $a_1$, $a_2$ and $a_3$ are ancilla states.}
	\label{fig:cloning1}
\end{figure}

In ref. \cite{ulrik2} a $1\rightarrow 2$ cloning map based on linear optics, homodyne detection and feedforward was proposed. A generalized version of this map is illustrated in fig.~\ref{fig:cloning1} and described in the figure caption. Setting the transmittivity to $T_2=1/2$ and the electronic gains to $g_x=g_p=\sqrt{2(1-T_1)/T_1}$ the input-output relation for one of the clones in the Heisenberg picture reads  
\begin{eqnarray}
\hat{a}_{clone1}=\frac{1}{\sqrt{2}}(\sqrt{\frac{1}{T_1}}\hat{a}_{in}+\sqrt{\frac{1}{T_1}-1}\hat{a}_{2}^\dagger+\hat{a}_{3}).
\label{input-output}
\end{eqnarray}
where $\hat{a}_{out}, \hat{a}_{in}, \hat{a}_{2}$ and $\hat{a}_3$ are the field operators associated with the output, the input and the ancilla vacuum fields respectively. The field operators are given by $\hat a=\hat x+i\hat p$ where $\hat x$ and  $\hat p$ are the amplitude and phase quadratures, respectively.
The map in eq.\ref{input-output} coincides with the one in ref.~\cite{ulrik2} for $T_1=1/2$ which was found to be the optimal Gaussian cloning map for completely unknown coherent states corresponding to a flat input alphabet. If on the other hand the number of possible input coherent states is finite the transformation in ref.~\cite{ulrik2} is no longer optimal. For a symmetric Gaussian distribution of coherent states with variance $V$:$ p(\alpha)=1/2\pi V\exp(-|\alpha|^2/2V)$,
the fidelity in (\ref{fidelity}) is   
\begin{equation}
\bar F=\frac{2 T_{1}}{2V(1-\sqrt{2T_{1}})^2+T_{1}+1}
\end{equation}
It is clear from this expression that the fidelity is a function of the knowledge of the input states through the variance, $V$, of the Gaussian alphabet. For a given variance, $V$, the maximized fidelity is 
\begin{equation}
\bar{F}=\left\{\begin{array}{l}
\frac{4V+2}{6V+1}, V\geq\frac{1}{2}+\sqrt{\frac{1}{2}}\\
\\
\frac{1}{(3-2\sqrt{2})V+1}, V\leq\frac{1}{2}+\sqrt{\frac{1}{2}}
\end{array}\right.
\label{optfid}
\end{equation}
which is obtained using the scheme in Fig.~\ref{fig:cloning1} with $T_1=\frac{1}{2}(1/(2V)+1)^2$ and $T_1=1$, corresponding to the upper and lower inequalities, respectively. These fidelities are identical to the ones found in ref.~\cite{cochrane04.pra} for optimized Gaussian cloning using an OPA. 
Let us consider these expressions in two extreme cases: If no a priori information is available about the distribution of coherent states, $V\rightarrow \infty$ and the fidelity averages to $\bar F=2/3$. In the other extreme, where complete information about the input state is at hand, $V=0$ and the fidelity is unity. In the following, we investigate experimentally the realistic intermediate regime where the width of the input distribution is finite and non-zero. 

We prepare the input coherent states by modulating a continuous wave laser beam (1064nm) at the frequency of 14.3~MHz. Two  electro-optical modulators inserted in the beam path were used to control the mean phase $ \langle p_{in}\rangle $ and amplitude $\langle x_{in} \rangle $ quadratures independently by separate low voltage function generators. Through the modulation, photons were transferred from the carrier into the sidebands, thus producing a pure coherent state at the modulation frequency. We set the modulation frequency to 14.3 MHz and defined the bandwidth of the coherent state to be 100 kHz.

\begin{figure}[h]
	\centering
		\includegraphics[width=8cm]{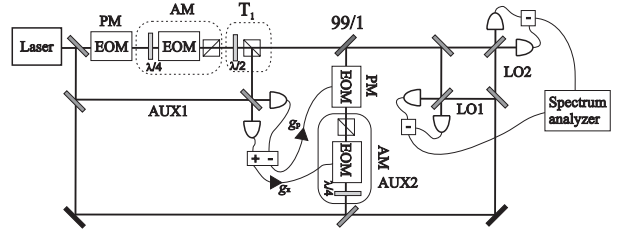}\caption{Experimental cloning setup. AM: Amplitude Modulator, PM: Phase Modulator, EOM: electro-optic modulator; LO: Local Oscillator; AUX: Auxiliary state; $T_{1}$: Variable beam splitter.}
	\label{fig:cloning}
\end{figure}

The pure coherent states are subsequently injected into the cloning machine. First the states are split into two parts using a variable beam splitter which is consisting of a half wave plate and a polarising beam splitter; thus any $T_1$ in eq.~(\ref{input-output}) is easily accessed by a simple phase plate rotation. The reflected part of the state is measured using heterodyne detection where $x$ and $p$ are simultaneously measured. This is done by interfering the signal with an auxiliary beam (AUX1) with a $\pi/2$ phase shift and balanced intensities; subsequently the two output are measured with high efficiency and low noise detectors, and the sum and the difference currents are constructed to provide a measure of $x$ and $p$. The outcomes are scaled with low noise electronic amplifiers and used to modulate the amplitude and phase of an auxiliary beam (AUX2), and subsequently combined by the remaining part of the signal employing a 99/1 beam splitter. This accomplishes a high efficiency conditional displacement operation. Finally, the displaced state is divided into two clones using a symmetric beam splitter and the two outputs are characterized using two homodyne detectors with intense local oscillator beams (LO1 and LO2). The signal power and variances of the input state as well as the output states are then measured using a spectrum analyzer with resolution bandwidth set at 100kHz  and video bandwidth at 300 Hz. Such measurements suffice to fully characterise the states due to the  Gaussian statistics of $x$ and $p$. Active electronic feedback loops were implemented at all interferences to ensure stable relative phases. From the power and variance measurements, we estimate the gain as well as the added noises associated with the cloning transformation. Using these values we calculate the fidelity for a given input alphabet using the expression 
\begin{equation}
\bar F=\frac{2}{\sqrt{(1+\Delta x^2+4V(1-\lambda_x)^2)(1+\Delta p^2+4V(1-\lambda_p)^2)}} 
\label{acloning}
\end{equation}
which is obtained from eq.~\ref{fidelity} by inserting an arbitrary Gaussian state (with variance $V$) in  replacement of $\Gamma(\alpha)$. $\lambda_x=\langle x_{clone}\rangle/\langle x_{in}\rangle$ and $\lambda_p=\langle p_{clone}\rangle/\langle p_{in}\rangle$ are the cloning amplitude gains.    
\begin{figure}[h]
	\centering
		\includegraphics[width=7cm]{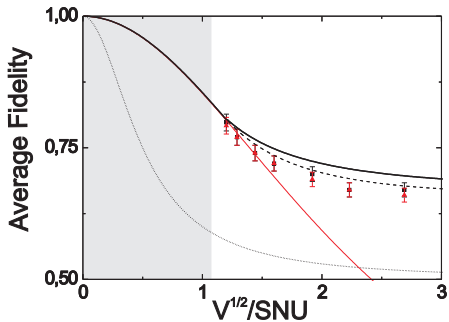}\caption{The average fidelity is plotted against  the width ($\sqrt{V}$) of the distribution of input states. The solid line corresponds to the theory and the red dot and black square correspond to average fidelities for clones 1 and 2. 
The dashed line takes into account that the amplifier used in the scheme was nonideal and some technical noise entered the cloning process. The dotted line corresponds to a measure and prepare strategy. The grey shaded area corresponds the region where the solution $T_1=1$ is optimal. The red line extends this solution into the region where it is not optimal.}
	\label{fig:cloningresults}
\end{figure} 

As an example, we consider a Gaussian input distribution with $V=1.72$ shot noise units (SNU). For this alphabet, the cloning machine is optimized by setting $T_{1}=0.83$ and $g_x= g_p=0.64$ corresponding to a cloning gain of $\lambda_x=\lambda_p=0.775$. We adjusted the beam splitter transmittance to this value and tuned the electronic gains to the optimized value while monitoring the optical gain of a test signal (through comparison between the input power and output power of the signal). For this specific experimental run, we measured an optical cloning gain of $0.775\pm0.005$ valid for all input states. After adjusting the gain to this value, the associated added noises in $x$ and $p$ were measured to $1.21\pm 0.02$ and $1.26\pm0.02$. Inserting these values in eqn.~(\ref{acloning}) we find an average cloning fidelity of $F=0.775\pm0.01$ which is arbitrarily close to the optimal value of $F=0.785$ (see eqn.~(\ref{optfid})). This experiment was repeated with different gains corresponding to different widths of the input alphabet and the results are summarized in Fig.~\ref{fig:cloningresults}. The solid curve in Fig.~\ref{fig:cloningresults} represents the ideal average fidelity given by eqn.~\ref{optfid}. Small deviations from ideal performance is caused by small inefficiencies of the heterodyne detector in the feedforward loop: The mode overlap between the auxiliary beam AUX1 and the signal beam was 99\% and the quantum efficiency of the associated detectors were 95\%$\pm 2\%$. Taking these parameters into account, the expected average fidelity follows the dashed curve which agrees well with the measured data. Note that all the measured data were corrected for the detection inefficiencies of the verifying detectors (amounting to 83\% and 85\%) to avoid an erroneous underestimation of the added noise and thus an overestimation of the fidelity.   
Note also that for $V<1/2+1/\sqrt{2}$ (corresponding to the gray shaded region in ~\ref{fig:cloningresults}), the best cloning strategy is a simple beam splitter operation, which is obtained in the present setup by setting $T=1$ and $g_x=g_p=0$, thus $\lambda_x=\lambda_p=1/\sqrt{2}$. In this case ideal performance is naturally achieved and the ideal solid curve and real dashed curve in Fig.~\ref{fig:cloningresults} are identical.  Since the detection efficiency is inferred out of the results, the actual measured performance will be only limited by the errors in estimating these efficiencies. 

We now proceed by considering another input alphabet. The coherent states are assumed to have a known and constant average phase but completely random amplitude. This input distribution was also considered theoretically in ref.~\cite{cochrane04.pra} and \cite{namiki06.xxx} where two different strategies were suggested for the experimental realizations. In the latter reference, however, the proposed strategy was not optimal and in the former reference the method relied on squeezing transformations to surpass the classical cloning strategy. Furthermore, the optimality of the suggested schemes were not proven in these references. In the following we show that the transformation depicted in Fig.~\ref{fig:cloning1} is optimal for special choices of the ancilla states $a_1$ and $a_3$, and the transmittances $T_1$ and $T_2$. We start by setting $T_1=1/2$ and $T_2=1$ and thus get the following transformation for one of the output clones
\begin{eqnarray}
x_{clone1}&=&x_{in}+\frac{1}{\sqrt{2}}x_3\\
p_{clone1}&=&\frac{1}{2}p_{in}-\frac{1}{2}p_1+\frac{1}{\sqrt{2}}p_3
\end{eqnarray}   
First assuming that the input ancillas($a_{1}$, $a_{2}$, and $a_{3}$) are vacuum states, the fidelity for this transformation is easily found using the expression~(\ref{fidelity}) and inserting a distribution with the above mentioned properties. We find $F=2/\sqrt{5}\approx 0.894$. This should be compared with the optimised measure and prepare strategy, which we conjecture to be associated with single quadrature detection followed by displacement of an optimally squeezed ancilla state in the quadrature direction corresponding the constant phase. The optimised squeezing factor is $\sqrt{1/2}$ of the undisplaced phase quadrature, and this measure and prepare strategy yields a fidelity of $F=2/\sqrt{3+\sqrt{2}}\approx0.828$~\cite{note}. Remarkably, our proposed scheme surpasses this value without the use of squeezed states. Although this cloning protocol surpasses the measure and prepare protocol, it is not the optimal Gaussian cloning machine for this input alphabet. If the input state $a_1$ is infinitely squeezed in the amplitude quadrature and $a_3$ is squeezed by a factor of $\sqrt{8/5}$, the cloning machine is optimal yielding a fidelity, $F=4(\sqrt{10}-1)/9\approx 0.961$. Hence, knowing the phase of the input coherent states, the cloning fidelity can be exceptionally high using a very simple scheme. 

Let us now prove the optimality of this scheme. A generic Gaussian cloning transformation is casted as
\begin{eqnarray}
x_{clone1}=\lambda_{x1}(x_{in}+n_{x1})\;\;\;\;\;\;p_{clone1}=\lambda_{p1}(p_{in}+n_{p1})\nonumber\\
x_{clone2}=\lambda_{x2}(x_{in}+n_{x2})\;\;\;\;\;\;p_{clone2}=\lambda_{p2}(p_{in}+n_{p2})\nonumber\\
\nonumber \end{eqnarray}
where $n_i$ are noise operators. Since the two clones are assumed to be identical we set $\lambda_x=\lambda_{x1}=\lambda_{x2}$ and $\lambda_p=\lambda_{p1}=\lambda_{p2}$ and because the amplitude of the input is completely random we must have $\lambda_x=1$. Furthermore by using the fact that the amplitude (phase) quadrature of clone 1 commutes with the phase (amplitude) quadrature of clone 2 we find $[n_{x1},n_{p2}]=[n_{x2},n_{p1}]=-2i$ which in turn yields the uncertainty relation 
\begin{equation}
\Delta^2n_{p} \Delta^2n_{x}\geq 1
\label{n}
\end{equation}
where $\Delta^2n_{x}=\Delta^2n_{x1}=\Delta^2n_{x2}$ and $\Delta^2n_{p}=\Delta^2n_{p1}=\Delta^2n_{p2}$. Now considering the commutation relation between conjugate quadratures of a single clone we find the uncertainty product $\Delta^2n_{x}\Delta^2n_{p}\geq|\frac{1-\lambda_p}{\lambda_p}|^2$ which is  minimized for $\lambda_p=1/2$. The minimum variances of the two output clones are therefore
\begin{eqnarray}
\Delta^2x=1+\Delta^2n_{x}\\
\Delta^2p=\frac{1}{4}(1+\Delta^2n_{p})
\end{eqnarray}   
By evaluating the Gaussian fidelity for these clones we find that it is maximized if the ancilla state is squeezed such that $\Delta^2n_p=\sqrt{5/2}$ and $\Delta^2n_x=\sqrt{2/5}$. The maximum fidelity is thus found to be $F=4(\sqrt{10}-1)/9$.

We now demonstrate cloning of coherent states with constant phases using the scheme in Fig.~\ref{fig:cloning1} with $a_1$ and $a_3$ in vacuum states. The experimental setup is slightly modified with respect to the one in Fig.~\ref{fig:cloning}. To enable direct detection of the amplitude quadrature in the feedforward loop, the auxiliary beam AUX1 is blocked and the sum of the currents produced in the two detectors is taken. This yields the amplitude quadrature and is correspondingly used to generate the amplitude displacement. The feedforward gain driving the phase displacement is set to zero, thus the phase quadrature is unaffected by the feedforward action.      
Since the amplitude quadrature of the input states is completely unknown, the electronic gain, $g_x$ is set such that the overall optical amplitude quadrature gain is unity. This maximises the average fidelity for this set of states. 
\begin{figure}[h]
	\centering
		\includegraphics[width=8cm]{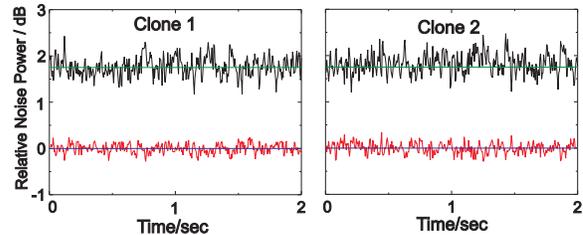}\caption{Spectral amplitude quadrature noise densities of the two clones (upper black traces) relative to the quantum noise level (lower red trace). The measurement was taken over a period of 2 seconds. The settings of the spectrum analyser were 14.3 MHz central frequency, 100 kHz bandwidth and 300 Hz video bandwidth. The added noise contributions are $1.8\pm0.1$dB and $1.85\pm 0.1$dB for clone 1 and clone 2 respectively. The optimal cloning limit (1.75 dB above the shot noise level) is pointed out by the solid green line.}
	\label{fig:amplitude cloning}
\end{figure} 
Finally, the clones are generated at the output of the third beam splitter. The verification procedure is the same as before, and a measurement run is depicted in Fig.~\ref{fig:amplitude cloning}. Making use of eqn.~(\ref{acloning}) we calculated the fidelity of the generated clones to be  $89.1\pm0.2\%$ and $88.7\pm0.2\% $. In this particular measurement run the gains for the amplitude quadratures were measured to be $\lambda_{x1}=0.98\pm0.01$ and $\lambda_{x2}=0.99\pm0.01$ for clone 1 and clone 2. The experimental cloning fidelity greatly exceed the classical fidelity of $82.8\%$ and is arbitrarily close to the optimal value of 89.4\% for non-squeezed ancillas. 


In conclusion, we have illustrated the intriguing relationship between cloning fidelity and prior partial information by proposing and experimentally demonstrating the state dependent cloning transformation of coherent states with superior fidelities. We found that the more prior information about the input states the greater is the cloning fidelities. This relationship is not only valid for cloning protocols, but also for other protocols such as teleportation and purification of quantum information. Since prior partial information is common in quantum information networks, we believe that the state-dependent cloning strategies presented in this paper as well as similar strategies for other protocols will have a vital role in future quantum informational systems.

We thank R. Filip and V. Josse for fruitful discussions. This work is supported by the EU projects COMPAS and SECOQC, and the danish research council (FTP).


\begin{thebibliography}{99}
\bibitem{note} If the vacuum state is displaced instead of the squeezed vacuum, the fidelity is $F=2/\sqrt{6}=0.817$.
\end{thebibliography}
\end{document}